# Curie Temperature for Small World Ising Systems of Different Dimensions

E.Z. Meilikhov, and R.M. Farzetdinova[*]

*RRC "Kurchatov Institute", 123182 Moscow, Russia*

**Abstract**

For Small World Ising systems of different dimensions, "concentration" dependencies $T_C(p)$ of the Curie temperature upon the fraction of long-range links have been derived on a basis of simple physical considerations. We have found $T_C(p) \propto 1/\ln|p|$ for 1D, $T_C(p) \propto p^{1/2}$ for 2D, and $T_C(p) \propto p^{2/3}$ for 3D.

## 1. Introduction

We consider the system of Ising spins in non-regular networks presenting the Small World – the graph with peculiar properties [1, 2]. "Ordinary" networks (regular and non-regular) refer to the lattices the sites of which are connected with their neighbours only. By contrast, in SW networks there are random connections both between near and far (in geometrical sense) sites. It is precisely these far, or long-range, links (shortcuts) that are responsible for the especial features of SW networks.

The Ising problem in SW networks arises in a similar way as for ordinary lattices – in network sites Ising spins are placed which interact (the interaction energy $J$) with their nearest neighbours only, that is with those ones which are directly connected with a given spin. However, in SW some geometrically removed spins turn out to be "the nearest neighbours" that, naturally, is favourable to magnetic ordering.

The existence or the absence of the ordered magnetic state is governed by the fraction $p$ of long-range links and the dependence of the interaction energy $J_{SW}$ ascribed to those links on their geometrical length. Numerical calculations for the power dependence $J_{SW} = J^+ r_{ij}^\alpha$ ($r_{ij}$ is the geometrical distance between sites $i$, $j$) show that the phase transition at a finite temperature is possible even in the one-dimensional system [3−7] though it occurs at $\alpha=0$ only, when the interaction energy along those links is independent of the distance and, certainly, at sufficiently high fraction of those links. At $\alpha>0$, the phase transition is absent [8].

Most conclusions about the properties of the Ising model in SW networks have been obtained by numerical Monte Carlo calculations. In the present paper, we derive analytical expressions for "concentration" dependencies $T_C(p)$ of the Curie temperature in SW-systems of different dimensions.

## 2. Thermodynamic theory for the concentration dependence of the Curie temperature

To gain some insight into the physical reason for the "concentration" dependence $T_C(p)$ (i.e., enhancing $T_C$ with increasing the "concentration" $p$ of long links in SW networks) let us consider arguments explaining the absence of the ordered ferromagnetic state in the one-dimensional system without long links and the possibility of that state appearing in a SW network. In the first case, splitting the spin chain into two domains with opposite magnetization results in increasing the system energy by $2J$. However, the boundary between those domains could be placed at any of $N$ chain sites that corresponds to raising the entropy by $k\ln N$. Thus, the variation of the free energy equal to $\Delta F=2J - kT\ln N$ is always negative at high enough number $N$, i.e., domain formation is efficient. In the case of SW networks, the possibility to choose the position of the domain boundary (without additional increasing the energy due to long links) is significantly limited: the number of sites "suitable" for that boundary lowers by about $pN \gg 1$ times comparing with the original site number $N$ and equals $\sim N/pN = 1/p$. Now, the variation of the free energy is $\Delta F \sim 2J - kT\ln(1/p)$. It is positive (i.e., domain formation is non-profitable) if $T < T_C^{(1D)}$ where

$$kT_C^{(1D)} \sim 2J/|\ln p|. \qquad (1)$$

---

[*] Corresponding author. Tel.: (095)-196-7684; fax: (095)-194-1994; e-mail: meilikhov@imp.kiae.ru (E. Meilikhov).

Thus, even the one-dimensional SW-system could be magnetically ordered. That conclusion agrees with the result of the exact solution for some particular one-dimensional SW-systems [4, 5, 9].

Another matter is the two-dimensional lattice where the creation of the domain with perimeter length of $L$ (in units equal to the lattice constant) leads to appearing $L$ pairs of spins of opposite directions at the domain's boundary that results in increasing the system energy by $2LJ$. To calculate the entropy associated with that boundary, one needs to estimate the number of ways to draw a closed boundary of length $L$. As in every site the boundary could "choose" one of three directions, the number of those ways is about $G = 3^L$ (as the boundary is closed, that number is somewhat overestimated but for large $L$ the error is insignificant [10]). Thus, the variation of the free energy equals $\Delta F \approx 2LJ - kT\ln G = L(2J - kT\ln 3)$. The ordered state is stable when that variation is positive, i.e., at $T < kT_C^{(2D)}$ where[1] $kT_C^{(2D)} = 2J/\ln 3 = 1.82J$. The presence of long links lowers the number of ways to draw the boundary: it could not travel through the sites possessing those links because the system energy would be higher. The number of those "forbidden" sites is on the order of $(pL^2)^{1/2} = p^{1/2}L$, and near those sites the boundary could choose not three but only two directions. Hence, the number of possible boundaries of length $L$ reduces to $G_p \sim 3^{L-p^{1/2}L} 2^{p^{1/2}L}$, and the variation of the free energy equal to $\Delta F \approx 2LJ - kT\ln G_p = L\{2J - kT[\ln 3 - p^{1/2}\ln(3/2)]\}$ is positive at $T < T_C^{(2D)} + \Delta T_C^{(2D)}$ where

$$\Delta T_C^{(2D)} / T_C^{(2D)} = \left[\frac{\ln 3}{\ln(3/2)\sqrt{p}} - 1\right]^{-1} \to 0.37 p^{1/2} \quad (p << 1). \qquad (2)$$

Analogous arguments also allow one to estimate the dependence $T_C(p)$ for three-dimensional case where creation of a three-dimensional domain with surface area of $S$ leads to the formation of $S$ spin pairs with opposite directions, resulting in increasing the energy by $2SJ$. If the number of variants to "extend" the surface in every site is $g \sim 1$, then the number of ways to create the domain with the surface area $S$ is about $G = g^S$. Thus, the variation of the free energy equals $\Delta F \approx 2S - kT\ln G = S(2J - kT\ln g)$. The ordered state is stable at $T < T_C^{(3D)}$ where $T_C^{(3D)} = 2J/\ln g$.

The exact value of the parameter $g$ is unknown, however it could be estimated through the following arguments. To prolong the unclosed surface consisting (in the cubic lattice) of the square facets one could attach one of the three new facets to every edge of the surface rim (border): one of them is parallel to the existing facet, and two other are perpendicular to it. However, it does not mean that $g=3$, since not every possible combination of facets, being attached to the adjacent edges, is acceptable. That is clarified by Fig. 1 where the piece of the border, consisting of two perpendicular edges, is displayed (the upper draft). Among $3^2=9$ possible combinations of new edges attached to those rim edges, four combinations indicated in Fig. 1 (lower drafts) are not acceptable. The other five combinations correspond to the effective value $g = 5^{1/2} \approx 2.25$. One gets even higher portion of unacceptable combinations considering the border piece consisting of three edges. If, for example, the latter is of the zigzag form, then for every former acceptable combination of two facets there are only two (instead of three!) suitable orientations of the third facet. That corresponds to ten ($5 \times 2 = 10$) possible variants of prolonging the border and to the effective value $g = 10^{1/3} \approx 2.15$. At increasing the number of elements of the zigzag border, $g \to \lim_{n \to \infty} (5 \cdot 2^{n-2})^{1/n} = 2$. (Analogous result occurs for the border of the meander form.) The number of acceptable variants diminishes further if one takes into account that the constructed surface has to be closed. Thus, one could consider $1 < g < 2$ and assume, for instance, $g = 1.5$. In doing so, one get $kT_C^{(3D)} = 2J/\ln g \approx 4.9J$, that agrees well with the almost exact solution $kT_C^{(3D)} \approx 4.5J$ [11] (to which value $g \approx 1.56$ corresponds).

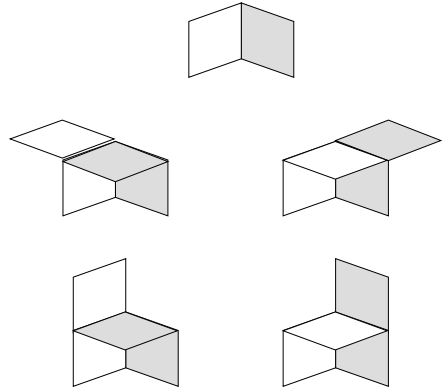

Fig. 1. The border piece consisting of two perpendicular edges (upper draft) and forbidden ways to prolong the surface.

---
[1] The Onsager exact solution $kT_C^{(2D)} = 2J/\ln(1+\sqrt{2}) = 2.27J$ changes this result unessentially.

The presence of long links diminishes the number of ways to extend the surface: it could not pass through the sites with those links as the system energy would be higher. The number of the "forbidden" sites is $\sim p^{2/3}S$, and near those the number of surface extension ways diminishes to $g - \Delta g$. Hence, the number of possible domains with surface $S$ lowers down to $G_p \sim g^{S-p^{2/3}S}(g-\Delta g)^{p^{2/3}S}$, and the variation of the free energy equal to $\Delta F \approx 2SJ - kT \ln G_p = S\{2J - kT[\ln g + p^{2/3}\ln(1-1/g)]\}$ is positive at $T < T_C^{(3D)} + \Delta T_C^{(3D)}$ where

$$\Delta T_C^{(3D)} / T_C^{(3D)} = \left[\frac{\ln g}{|\ln(1-\Delta g/g)| p^{2/3}} - 1\right]^{-1} \approx$$
$$\approx \frac{|\ln(1-\Delta g/g)|}{\ln g} p^{2/3} \qquad (3)$$

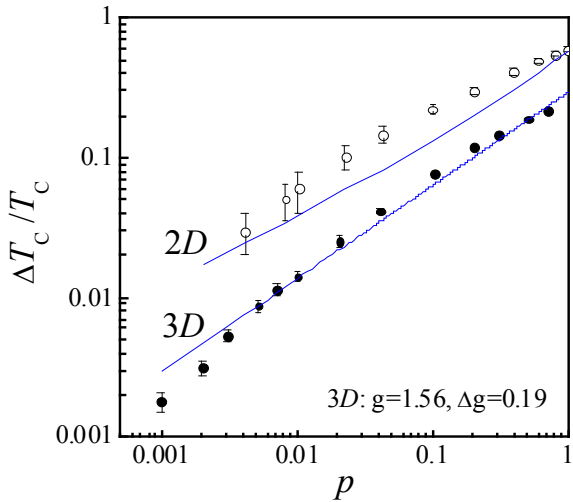

Fig. 2. "Experimental" [6] (points) and theoretical (solid curves) dependencies $\Delta T_C(p)$ for 2D- and 3D-networks.

The value $\Delta g$ has to be not too large since in the three-dimensional space the surface has many more opportunities to round the "forbidden" site. Therefore, one could think that $\Delta g \ll 1$.

The dependencies (1) − (3) are supported by numerical calculations [4, 6]. In Fig. 2, "experimental" results obtained in the course of such calculations [6] and theoretical ( Eqs. (2), (3) ) dependencies $\Delta T_C(p)$ for two- and three-dimensional networks are shown, where no fitting parameters are not used for the 2D-network. As for the 3D-network, magnitudes $g = 1.56$ (providing the exact $T_C^{(3D)}$ value) and $\Delta g = 0.19$ have been accepted; then $\Delta T_C^{(3D)} \approx 0.29 p^{2/3}$. In spite of approximate nature of the theoretical model that explains increasing $T_C$ with $p$ by lowering the system entropy, there is qualitative and quantitative agreement with the numerical experiments.

## Acknowledgements

This work was supported by Grants Nos. 03-02-17029, and 04-02-16158 of the Russian Foundation of Basic Researches.